\documentclass[12pt]{article}
\usepackage{epsfig}

\newcommand{\bec}[1] {\begin{equation}\label{#1} }
\newcommand{\eec} {\end{equation} }

\newcommand{\beq}{\begin{equation} }
\newcommand{\eeq}{\end{equation}}

\newcommand{\bea}{\begin{eqnarray}}
\newcommand{\eea}{\end{eqnarray}}

\newcommand{\at}{\bar{t}}
\newcommand{\ab}{\bar{b}}

\newcommand{\la}{\mbox{$\lambda$}}

\begin{document}

\vspace*{-0.5cm}
\begin{flushright}
UR-1646\\
OSU-HEP-01-13
\end{flushright}
\vspace{0.5cm}

\begin{center}
{\Large
{\bf QCD Loop Corrections to Top Production and Decay at
$e^+ e^-$ Colliders  } }

\vspace*{1.cm}
 Cosmin Macesanu 
\footnote{Email address:  mcos@pas.rochester.edu}

\vspace*{0.5cm}
{\it Department of Physics and Astronomy, University of Rochester\\
Rochester, NY~14627-0171, USA\\}
{\it Department of Physics, Oklahoma State University
\footnote{Current address.}\\
Stillwater, OK~74078, USA\\}
\end{center}

\begin{abstract}

We present a computation of QCD next-to-leading order virtual corrections
to the top production and decay process at linear colliders. The 
top quarks are allowed to be off-shell and
the production and decay subprocesses are treated together, thus
allowing for interference effects.  The framework employed for our computation
is the double pole approximation (DPA). We describe the implementation of
this approximation for the top production and decay process
and compare it with the implementation of DPA for the evaluation
of QED corrections to the $W$ pair production at LEP II. Similarities
and differences between the two cases are pointed out. One result
of interest is the incomplete cancelation of interference corrections.
Other results include
values for the total NLO top production cross section, and the impact
the nonfactorizable (interference) corrections have on the top invariant
mass distribution.

\end{abstract}

\section{Introduction}

The top quark might prove to be one of the most interesting elementary particles discovered so far. Since its mass is so large (close
to the electroweak symmetry breaking scale), it can be assumed
 that study of top properties will allow us to learn new things about the
physics at the next energy scale. So far, though, 
due to low statistics in the Run I at Tevatron, our knowledge
of the top quark is rather limited.
Future runs at the Tevatron and the LHC will provide more information, 
but the analysis of this new data 
will be complicated by large QCD backgrounds. 

 An $e^+ e^-$ collider with center-of-mass energy at and above the top 
threshold promises to provide a clean environment in which to perform
precision studies of the top quark \cite{orange}.
 Indeed,
it is conceivable that at such a machine the study of top can be performed
with a precision similar to that 
aquired in the study the $W$ boson at LEP II.
This means order \%  (and better) measurements of 
differential cross sections for processes involving the top quark. 
Such a precision in measuring experimental quantities implies the need
for a like precision in our theoretical understanding of these processes.
This in turn requires 
the inclusion of radiative corrections in our 
predictions.

A wealth of information about the top quark (like its mass, width, 
strong and Yukawa coupling constants) can be extracted from 
measurements performed at the production threshold.
For this region, comprehensive theoretical
studies (NNLO computation with resummation of large logarithms,
careful treatment of the renormalon ambiguity) have already been performed
(\cite{top_thres} and references therein). Going to higher energies, 
we can study the V-A structure of the top quark 
couplings to the gauge bosons ($ \gamma, Z$ and $W$) \cite{anom_coup}.
The information about couplings can be obtained by using spin correlations:
the top quarks are produced in certain spin states, as dictated by the
top - $ \gamma, Z$ couplings. Since the top decays before hadronization,
the spin states of the top directly influence the angular 
distributions of its decay products. Simulations show that by analyzing
kinematic variables of final state particles it is possible to measure
the top-gauge boson couplings at the several percent level \cite{orange}.

The subject of this paper is the evaluation of
NLO QCD corrections to the top production and decay process above threshold
at a Linear Collider. 
 Previous work in this area includes the study of corrections
to the top production subprocess, with virtual and soft gluons
 \cite{jersak} 
as well as with hard gluon radiation
(\cite{korner1}, \cite{parke1}, \cite{arndt1} are just some examples),
and to the top decay subprocess with virtual and
hard gluons together (\cite{jeza1}, \cite{andrej}, \cite{oakes}).

Using these results, one can try to understand the top production and
decay process by assuming that the intermediate top quarks
are on-shell (narrow
width approximation)  and treating the subprocesses separately
\cite{schmidt}. This assumption is usually reasonable;
the result for the total cross section is valid up to terms of 
order $\Gamma_t/m_t \sim 1 \%$. 
However, the effects of finite top width can be
important in some differential cross sections; and if precision
better than percent level is needed, the production and decay processes
have to be considered  together,
by allowing the top quark to go off-shell.
These corrections can be thought as being due to interference between
production and decay, and are also known as nonfactorizable corrections.
Some results concerning the nonfactorizable corrections for the top
production process have already been published \cite{bbc_top}; 
we present here a more complete analysis.

 We shall perform this computation and present the results at the parton 
level only.  
We assume that the issues related to jet reconstruction and 
identification can been solved, and the final state contains two $W$ bosons
\footnote{ At the Monte Carlo level, we actually allow the on-shell $W$'s
to decay, either semileptonically or into a pair of massless quarks, but 
in the latter case we do not take 
into consideration QCD corrections to the W decay.}
and two $b$ quarks (the real gluon radiation case has been analyzed in 
\cite{hardglu}, \cite{mrst98}).
Even at this level, the complete computation of all the diagrams contributing
to this final state (Born and next to leading order) is a very difficult
task. Therefore, we shall employ the double pole approximation (DPA) which
means taking into account only the diagrams which contain a top - antitop pair.

Here, it is worth mentioning the strong similarities 
between the evaluation of QCD corrections to the process of interest to us:
\bec{tree_p}
 e^+ e^- \rightarrow t\ \bar{t}\ \rightarrow b\ W^+\ \bar{b}\ W^- 
\eec
and the evaluation of QED corrections to the $W$ pair production and 
decay process ($e^+ e^- \rightarrow W^+\ W^- \rightarrow 4f) $ at LEP.
The issues which arise in the two computations are similar,
because in both cases we are dealing with the production and
decay of heavy unstable particles.
Our treatment is largely similar to the one used for the electroweak 
process \cite{racoon}, \cite{ddr}, \cite{BBC}. But there are
some differences, both in the implementation of the DPA approximation
and  in the number and type of terms which contribute to the final result
(the latter being due to the fact that in our case the
intermediate off-shell particles are fermions, and not bosons).  
These differences will be pointed out in the course of our discussion.

 The outline of the paper is as follows. In sect. 2 we lay out in some detail
the theoretical framework in which we perform our computation.
This includes a short description of the DPA method, with examples 
of evaluation of NLO amplitudes in this approximation. The main point 
of this section is that the results for the amplitudes
corresponding to interference diagrams are similar to results 
previously obtained
for the $W$ pair production process, while for the vertex corrections and
fermion self-energy diagrams there are differences between the two cases.
In order to facilitate comparisons with the on-shell approach, we also
formulate our results in terms of corrections to the production and 
decay subprocesses and interference contributions. The gauge invariance
of the total and partial amplitudes in DPA is manifest in this formulation.
Sect. 3 contains some details on the design of the Fortran code we used
to obtain our numerical results, which are presented in sect. 5.
 A short discussion of a slighty
different approach in evaluating the non-factorizable corrections 
(used in \cite{bbc_top})
is presented in sect. 4. We end with the conclusions.


\section{Amplitudes in DPA}

\begin{figure}[ht!] 
\centerline{\epsfig{file=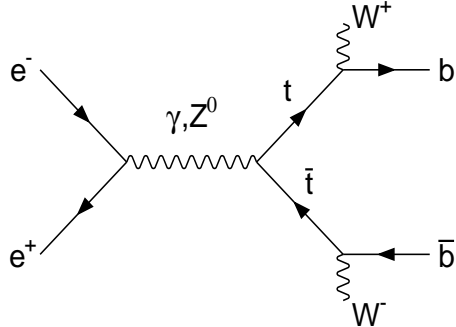,height=2.in,width=2.5in}}
\caption{The top-antitop diagrams contributing to the process (\ref{proc1}).}
\label{tree_level}
\end{figure}

\begin{figure}[t!] 
\centerline{\epsfig{file=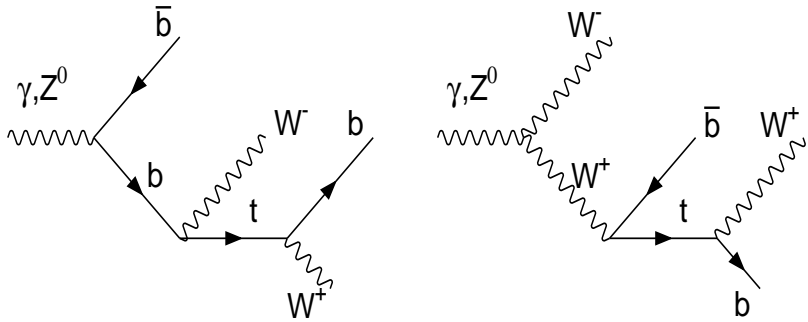,height=2.in,width=4.5in}}
\caption{Single top diagrams contributing to the process (\ref{proc1}).}
\label{sin_res_diag}
\end{figure}

Since it decays very fast, the top quark is not studied directly,
but through its decay products. 
In top pair production at linear colliders,
the relevant process is: 
\bec{proc1}
 e^+ e^- \rightarrow\ b\ W^+\ \bar{b}\ W^- 
\eec
At tree level, the diagrams contributing to this process can be split
into 3 classes: diagrams which contain a top-antitop
pair (Fig. \ref{tree_level}), diagrams which contain either one top quark
or a top antiquark (Fig. \ref{sin_res_diag} and charge conjugates),
 and diagrams which 
do not contain any top (there are about 50 such diagrams).  
The evaluation of all these amplitudes can be performed using 
 one of the automated tree level amplitude computation programs,
like MADGRAPH (\cite{madgraph}, see also \cite{treetop}).
 The computation of QCD corrections to all
tree level diagrams, however, increases the
 degree of complexity by quite a lot, and is probably not feasible yet.

Hence the need to use some methods which will simplify 
the computation, but will provide approximate results for the 
virtual corrections. One such method is the {\it double pole approximation} 
(DPA), which makes use of the fact that, looking
for top pair production, we are interested in a specific region
of the final state phase space of (\ref{proc1}). This region 
is defined by the requirement that the invariant mass of the $W\ b$
pair is close to the top mass: 
$ p_t^2,\ p_{\bar{t}}^2  \approx m_t^2$, where
$ p_t = p_{W^+} + p_b  $ and $ p_{\bar{t}} =  p_{W^-} + p_{\bar{b}} $.
In this region, the amplitudes corresponding to the top-antitop diagrams
are enhanced by the two resonant propagators coming from the two intermediate
top quarks\footnote{We
use the notation $\bar{m}_t^2 = m_t^2 - i m_t \Gamma_t$}: 
\bec{fdask} {\cal{M}} \sim  \frac{1}{p_t^2 - \bar{m}_t^2}\ 
\frac{1}{p_{\bar{t}}^2 - \bar{m}_t^2}\ 
\eec
(the diagrams in (Fig. \ref{tree_level}) are therefore called
doubly resonant) and we can neglect the contributions coming from
the singly resonant diagrams (Fig. \ref{sin_res_diag}) or nonresonant ones,
which are {\it reduced}
by factors of ($\Gamma_t/m_t$) or $(\Gamma_t/m_t)^2$
 with respect to the doubly resonant contributions.

\begin{figure}[tb] 
\centerline{\epsfig{file=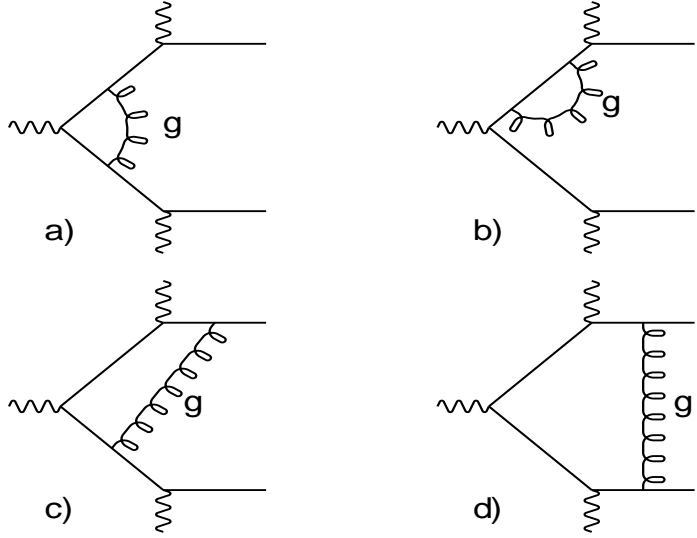,height=8.cm,width=11cm}}
\caption{Feynman diagrams for virtual gluon corrections to top production and decay.}
\label{vir_diag}
\end{figure}

The real gluon corrections to the doubly resonant diagrams have been computed
in \cite{hardglu}.
The aim of this paper is the DPA evaluation of the virtual corrections
to the top production and decay process.
Some of Feynman diagrams contributing to these corrections
are presented in Figure \ref{vir_diag}. These diagrams can be roughly 
divided into
two classes : corrections to particular subprocesses -- the vertex and 
fermion self-energy diagrams in Fig. \ref{vir_diag} a) and b) 
respectively -- and 
interference type corrections (Fig. \ref{vir_diag} c) and d)). Strictly
speaking, the vertex and self-energy diagrams also contribute to interference
between subprocesses; but, for the sake of brevity,
we shall refer to the diagram in Fig. \ref{vir_diag} a) as the production
vertex correction diagram, and so on. Also, in the following, we will denote
the tree level amplitude (Fig. \ref{tree_level}) by ${\cal{M}}^0$, and
the amplitude for the first order virtual corrections by ${\cal{M}}^{vg}$:
\bec{mvg_amp}
 {\cal{M}}^{vg} = {\cal{M}}_{t\bar{t}} + {\cal{M}}_{t b} +
  {\cal{M}}_{\bar{t}\bar{b}} + 
{\cal{M}}_{b\bar{t}} + {\cal{M}}_{t\bar{b}} + {\cal{M}}_{b\bar{b}} 
\eec 
Here, the  first three terms correspond to the three off-shell vertex
corrections (which include in a suitable way the fermion self-energies,
as described in section 2.4), 
and the last three terms come from the interference diagrams.

\subsection{Interference diagrams}

We start by discussing the evaluation of interference diagrams in 
Figs. 3c), 3d). Technically, this is one of the more involved  
computations we have to perform. The full evaluation of the 
$\at - b$ interference amplitude\footnote {The
Feynman gauge is used for the gluon propagator.}:
{\small
$$ {\cal{M}}_{b\bar{t}} =
 \bar{u}(b) \left[ (-i g_s^2)
 \int \frac{d^4 k}{2\pi^4}\ \frac{1}{k^2 + i\epsilon}\
\gamma^\mu \ \frac{\not{p}_b-\not{k} + m_b}{(p_b-k)^2 - m_b^2}\
\not{\epsilon}_{W^+} \ \frac{\not{p}_t-\not{k} + m_t}{(p_t-k)^2 - \bar{m}_t^2}\
{\Gamma}_{\gamma,Z_0}\ \right.
$$
\bec{mbbart}
\left. \times\
\frac{-\not{p}_{\bar{t}} - \not{k} + m_t}{(p_{\bar{t}}+k)^2 - \bar{m}_t^2}\
\gamma_\mu \right] \ 
\frac{-\not{p}_{\bar{t}} + m_t}{p_{\bar{t}}^2 - \bar{m}_t^2}\
\not{\epsilon}_{W^-} \ v(\bar{b})
\eec }
is, for example, quite a difficult task.
However, the only terms of interest to us in DPA are those which
have resonances at the top and antitop quark propagator poles. This
simplifies our task greatly. 
The doubly resonant terms can be extracted with the help of the following
observation: if the virtual gluon in the loop is hard, then 
the quantity in  brackets does not have any singularity, and the overall
resonant structure for this diagram is given only by the pole due to 
the antitop propagator:
$ {\cal{M}}_{b\bar{t}}(\hbox{hard gluon}) \propto 1/(p_{\bar{t}}^2 - \bar{m}_t^2)$.
This means that any doubly resonant terms contribution to
$ {\cal{M}}_{b\bar{t}}$ are entirely due to soft 
virtual gluons. 
Therefore, we can neglect the $\not{k}$ terms in 
the numerator of (\ref{mbbart}).
Following \cite{ddr}, we shall call this approximation 
the {\it extended soft gluon approximation} (ESGA)
\footnote{In the standard soft gluon approximation, $k^2$
terms in the denominator of top quark propagators would also
be neglected; we do not do this here for computational reasons
 (see also \cite{ddr}, \cite{kos}).}.

 With the help of the transformations:
\bec{trans_1}
\gamma^\mu \ (\not{p}_b + m_b) = ( -\not{p}_b + m_b)\ \gamma^\mu + 2p_b^\mu
\rightarrow 2p_b^\mu
\eec
$$ 
(-\not{p}_{\bar{t}} + m_t)\ \gamma_\mu = \gamma_\mu \ (\not{p}_{\bar{t}} + m_t)
- 2 p_{\bar{t} \mu} \rightarrow - 2 p_{\bar{t} \mu}
$$
(the term  $(\not{p}_{\bar{t}} + m_t)$ 
on the second line is neglected, since 
it would lead to a singly resonant contribution),
the following result is  obtained for the amplitude (\ref{mbbart}):
$$ {\cal{M}}_{b\bar{t}}(DPA+ESGA) = 
\frac{\alpha_s}{4\pi}\
 {\cal{M}}_0 \ \times \ (-4 p_b p_{\bar{t}})( p_t^2-\bar{m}_t^2 ) 
$$
\bec{mbbtr}
\times\
\int \frac{d^4 k}{i\pi^2}\ \frac{1}{k^2+ i\epsilon}\ \frac{1}{k^2-2kp_b}\  
\frac{1}{(p_t-k)^2 - \bar{m}_t^2}\ \frac{1}{(p_{\bar{t}}+k)^2 - \bar{m}_t^2}\
\eec
The result is proportional to the tree level amplitude --
 in the DPA, the virtual corrections 
due to interference factorize.
The proportionality factor includes the scalar 4-point function
(the integral on the second line of Eq. (\ref{mbbtr})) 
${\cal{D}}^0_{b\bar{t}} = {\cal{D}}^0(-p_b,-p_t,p_{\bar{t}},0,m_b,\bar{m}_t,\bar{m}_t)$.
\footnote{
 For the scalar one-loop integrals appearing here
 we use the following notation:
$$ {\cal{D}}^0(p_1,p_2,p_3,m_0,m_1,m_2,m_3) = 
\int \frac{d^4 k}{i \pi^2}\ \frac{1}{N_0\ N_1\ N_2\ N_3}\ 
$$
$$ {\cal{E}}^0(p_1,p_2,p_3,p_4,m_0,m_1,m_2,m_3,m_4) = 
\int \frac{d^4 k}{i \pi^2}\ \frac{1}{N_0\ N_1\ N_2\ N_3\ N_4}\ 
$$
with the denominators :
$$ N_0 = k^2 - m_0^2 + i\epsilon, \hbox{~~~~~} 
N_i = (k+p_i)^2 - m_i^2 + i\epsilon, \hbox{~~~~~} i=1,...,4
$$
}

Let's discuss shortly the resonant behavior of the DPA amplitude
in (\ref{mbbtr}). 
Apparently, the result for ${\cal{M}}_{b\bar{t}}$ has a single pole at 
$p_{\bar{t}}^2 = m_t^2$ (the other pole  being canceled 
by the multiplicative term $ p_t^2-m_t^2 $).
 However, if the
top (or antitop) goes on-shell, the ${\cal{D}}^0$ function 
acquires an infrared singularity (in the zero top width limit;
this singularity is regularized by the top width). 
Since the infrared singular type terms have 
a logarithmic structure (this can also be reasoned from power 
counting arguments), this indicates that ${\cal{D}}^0$ has the 
following behavior close to the top resonances:
\bec{res_beh}
 {\cal{D}}^0_{b\bar{t}} \sim a_1 \hbox{log}(p_t^2 - \bar{m}_t^2) + 
a_2 \hbox{log}(p_{\bar{t}}^2 - \bar{m}_t^2) 
\eec
Here $a_1$ and $a_2$ are terms which are finite when {\it either}
the top or antitop quark go on-shell.
Formally, then, the overall resonant behavior of the 
interference amplitude $ {\cal{M}}_{b\bar{t}}$
in DPA is of type $pole \times logarithm$: 
$$ {\cal{M}}_{b\bar{t}} \sim \tilde{{\cal{M}}}_0\ \hbox{log}(p_t^2 - \bar{m}_t^2) \ 
\frac{1}{p_{\bar{t}}^2 - \bar{m}_t^2} $$
rather than $pole \times pole$,
as it is for the tree level amplitude or for the 
corrections to production or decay subprocesses.

%

Using the same techniques, similar results are easily obtained for the
other two interference diagrams.
In the soft gluon approximation (and DPA):
\bec{mtbbr}
{\cal{M}}_{t\ab}(DPA+ESGA) = 
\frac{\alpha_s C_F}{4\pi}\
 {\cal{M}}_0 \ \times \ (-4 p_t p_{\ab})( p_{\at}^2-\bar{m}_t^2 ) \
 {\cal{D}}^0_{t\ab}
\eec
\bec{mbbbr}
 {\cal{M}}_{b\bar{b}}(DPA+ESGA) = 
\frac{\alpha_s C_F}{4\pi}\ {\cal{M}}_0 \ \times \ (-4 p_b p_{\bar{b}})( p_t^2-\bar{m}_t^2 )
( p_{\bar{t}}^2-\bar{m}_t^2 ) \ {\cal{E}}^0_{b\ab}
\eec
where ${\cal{D}}^0_{t\ab} = {\cal{D}}^0 (-p_{\ab},-p_{\at},p_t,0,m_b,\bar{m}_t,\bar{m}_t)$ and
\bec{e0}
{\cal{E}}^0_{b\ab} = {\cal{E}}^0
(-p_b,-p_t,p_{\bar{t}},p_{\bar{b}},\mu,m_b,\bar{m}_t,\bar{m}_t,m_b)
\eec
is the scalar 5-point function (here $\mu$ is the infinitesimally small
gluon mass  needed for the regularization of infrared divergent
behavior of ${\cal{E}}^0_{b\ab}$).


We end this section with some comments on the numerical magnitude
of interference terms. Since the resonant behavior of these terms is
of $pole \times log$ type, it might be expected that they are less important
numerically that the double pole terms. However, 
analytic expressions for the ${\cal{D}}_0$ function (\cite{ddr},
\cite{mel_yak}, \cite{BBC})
show that, although the 
coefficients $ a_1, a_2$ in (\ref{res_beh}) are finite when one of 
the top or antitop quark goes on shell, they will diverge when both particles
go on-shell simultaneously:
$$ a_i\ \sim \ 
\frac{1}{c_{1i} (p_t^2 - \bar{m}_t^2) + c_{2i} (p_{\bar{t}}^2 - \bar{m}_t^2)}
$$
Therefore, the leading logarithms in the scalar 4 and 5-point functions
will be enhanced
by factors of order $m_t/\Gamma_t$ near the top, antitop quark mass resonances.


\subsection{Vertex corrections}

\begin{figure}[t!] 
\centerline{\epsfig{file=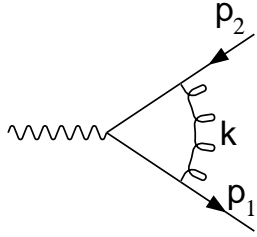,height=1.6in,width=1.6in}}
\caption{General vertex correction diagram.}
\label{ver_cor}
\end{figure}

 The results for the interference diagrams
are completely analogous to
results obtained in the $W$ pair production computation.
However, for the off-shell vertex and self-energy
corrections diagrams, the results in the top case are different.
Consider for example, the amplitude for the general vertex correction
in Figure \ref{ver_cor}: 
$$ \delta \Gamma^{\mu} = \frac{\alpha_s}{4 \pi} \
\int \frac{d^4 k}{i \pi^2}\ \frac{1}{k^2}\ \gamma^{\nu}\
\frac{\not{p}_1 - \not{k} + m_1}{(p_1 - k)^2 - \bar{m}_1^2}\
\gamma^{\mu}(C_V + C_A \gamma^5 ) \
\frac{-\not{p}_2 - \not{k} + m_2}{(p_2 + k)^2 - \bar{m}_2^2}\
\gamma_{\nu}
$$
Upon evaluation (and keeping only the vector part)
the result can be written in terms of eight form factors, each
of them multiplying a different tensor quantity:
$$
\delta \Gamma^{\mu}_V = \frac{\alpha_s}{4 \pi} C_V\
\left[ {\gamma^{\mu}} {F_2} + { (\not{p}_1 - m_1)\gamma^{\mu}} {F_4}
 + {\gamma^{\mu}(-\not{p}_2 - m_2)} {F_6} 
\right.
$$
\bec{vertex}
 \left. + {(\not{p}_1 - m_1)\gamma^{\mu}(-\not{p}_2 - m_2)} {F_8} + 
{p_1^{\mu}} {F_1} +  
(\not{p}_1 - m_1) p_1^{\mu} F_3   +\ldots \right]
\eec
(expressions for the scalar form factors $F_1,\ldots, F_8$ can be found
in the Appendix).
In the on-shell case, only the $F_2$ (electric dipole) and $F_1$ (magnetic dipole momentum) form factors contribute. It might be expected that in the 
double pole approximation we can drop the other terms, too, since
they  have a zero at 
$\not{p}_1 = m_1$ (or $\not{p}_2 = -m_2$) which will
cancel one pole (or both) in the amplitude. 
However, the 
form factors themselves
may have a resonant structure when the  particles go on shell.

Consider the top decay vertex correction. In this case,
$p_1 \rightarrow p_b , p_2 \rightarrow -p_t$, and only four terms survive
in Eq. (\ref{vertex}); the corresponding form factors
contain terms which are proportional to the scalar 3-point function:
$$ F_i \ \sim \ C^0_{tb} \ =
\int \frac{d^4 k}{i \pi^2}\ \frac{1}{k^2}\ 
\frac{1}{(p_b - k)^2 - m_b^2}\
\frac{1}{(p_t - k)^2 - \bar{m}_t^2}\
\hbox{~~~~~~~} i= 1,2, 5,6 $$
which has 
a logarithmic resonant behavior:
$$C^0_{tb} \ \sim \ \hbox{log}(p_t^2 - \bar{m_t}^2).$$ 
Therefore, the contribution 
of $i=5,6$ terms to the top decay vertex correction is doubly
resonant, although of type $pole \times log$ rather than double pole.
Because these logarithms are not multiplied by large factors (as in the
case of the interference diagrams), these terms can be expected to be
numerically small; for consistency reasons it is still desirable to 
include them in the final result.

Similar results are obtained for the correction to the antitop decay 
vertex (we keep the $i=1,2,3,4$ terms in this case). In the case of the 
$t \ - \ \bar{t}$ vertex, though, both fermions are off-shell; as a consequence,
there are no resonant logarithms when either the top or antitop 
quark goes on-shell, and we keep only the $i=1,2$ terms.
 
 It follows that in the general expression (\ref{vertex})
it is necessary to keep the terms which contain $F_1$ to $F_6$  
(the $F_7$ and $F_8$ terms can be dropped),
and we don't have factorization for the interference part anymore. This is 
different from what happens in the $W$ pair production process,
where the DPA factorization holds even for the vertex corrections. This
difference between the two cases is due to the fact that in one 
process the intermediate particles
are fermions, while in the other they are bosons.


\subsection{Renormalization and fermion self-energy}

\begin{figure}[t!] 
\centerline{\epsfig{file=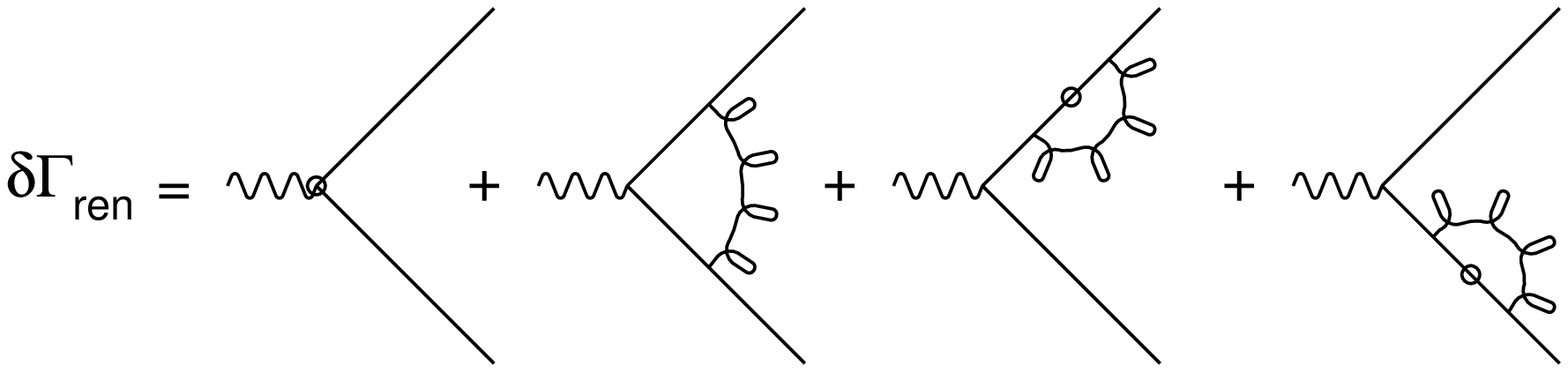,height=1.4in,width=5.in}}
\caption{Terms contributing to the renormalized vertex; the dots
 represent counterterm insertions.}
\label{ren_vert}
\end{figure}

For the renormalization of the ultraviolet divergences appearing
in the evaluation of the vertex and fermion self-energy corrections
we use the counterterm method. What
this amounts to is replacing the bare vertex 
correction in Fig. \ref{ver_cor} (which is UV divergent) with a 
finite renormalized vertex correction (Fig. \ref{ren_vert}):
\bec{ver_ren}
\delta \Gamma^{\mu}_{ren} \ = \
\delta \Gamma^{\mu} + \Gamma^{\mu} \delta Z_2 + 
\frac{1}{2}(-i  
\hat{\Sigma}_2(\not{p}_1) ) \frac{i}{\not{p}_1 - m_1} \Gamma^{\mu}
+ 
\frac{1}{2}\Gamma^{\mu}\frac{i}{-\not{p}_2 - m_2} (-i  
\hat{\Sigma}_2 ) (\not{p}_2)
\eec
which also includes the contributions of the fermion self-energy
diagram.

 The first two terms in Eq. (\ref{ver_ren}) are what is usually defined
as the renormalized vertex. The last two terms are one half of the 
renormalized fermion and antifermion self-energy:
\bec{ren_se}
  \hat{\Sigma}_2(\not{p}) = \Sigma_2(\not{p}) - 
(\pm \not{p} - m) \delta Z_2 -\Delta m
\eec
where $ \delta Z_2$ and $ \Delta m$ are coefficients of the counterterms
in the Lagrangian density 
(the $\pm$ sign corresponds to the particle/antiparticle case).
In the above 
equation, $\Sigma_2$ stands for the {\it bare} fermion self-energy:
\bec{bare_fse}
 \Sigma_2(\not{p}) = \frac{\alpha_s}{4 \pi} \
\int \frac{d^4 k}{i \pi^2}\ \frac{1}{k^2}\ \gamma^{\mu}\
\frac{\not{p} -\not{k} + m }{(p-k)^2 - \bar{m}^2}\ \gamma_{\mu}
\eec
Upon evaluation, the result for Eq. (\ref{bare_fse}) can be 
separated into a scalar and a spinorial component:
$$ \Sigma(\not{p}) = (\pm \not{p} - m) \Sigma_a(p^2) \ + \ m \Sigma_b(p^2)$$
With these
notations, the renormalized vertex correction can be written as:
\bec{ren_ver2}
\delta \Gamma^{\mu}_{ren} \ = \
\delta \Gamma^{\mu} \ + \ \frac{1}{2}\ \Delta Z_2(p_1)\ \Gamma^{\mu}
\ + \ \frac{1}{2}\ \Gamma^{\mu}\ \Delta Z_2(p_2) 
\eec
with
\bec{del_Z2}
 \Delta Z_2(p)\ = \ \Sigma_a(p^2) + 
\frac{ m \Sigma_b(p^2) - \Delta m}{\pm \not{p} - m}
\eec
The counterterm coefficient $\Delta m$ is fixed by the 
on-shell renormalization condition: 
\bec{del_m}
 \hat{\Sigma}_2(\not{p}= m) \ = \ 0 \ \Rightarrow \ 
\Delta m =  m \Sigma_b(m^2)
\eec
Also, in the on-shell limit,
\bec{del_Z2_os}
 \Delta Z_2(p) \arrowvert_{\not{p}\rightarrow m}
= \left( \Sigma_a(p^2) + \left.
\frac{ \partial \Sigma_b(\not{p}\cdot\not{p})}{\partial\not{p}} 
\right\arrowvert_{\not{p}\rightarrow m}
 \right) =  \left. \frac{ \partial \Sigma_2(\not{p})}{\partial\not{p}} 
\right\arrowvert_{\not{p}\rightarrow m} = \delta Z_2
\eec
where for the last equality we have used the renormalization
condition
$$  \left.
\frac{ \partial \hat{\Sigma_2}(\not{p})}{\partial\not{p}} 
\right\arrowvert_{\not{p}\rightarrow m}
 = 0
$$

It is convenient to write the contribution of the fermion self-energy
diagrams in a form similar to that of Eq. (\ref{vertex}). Using the resummed
top quark propagator:
\bec{top_prop}
\frac{i}{\pm \not{p} - m} \ \longrightarrow \
\frac{i(\pm \not{p} + m)}{ p^2 - \bar{m}^2}
\eec
we obtain the following result for $\Delta Z_2$ :
\bec{del_Z2_fin}
\Delta Z_2(p) =  \left( [\Sigma_a(p^2) + 2 \Sigma_{ir}(p^2)] \ + \
\frac{\Sigma_{ir}(p^2)}{m}( \pm \not{p} - m ) \right)
\eec
with 
$$  \Sigma_{ir}(p^2) = 
m \frac{ \Sigma_b(p^2) - \Sigma_b(m^2)}{p^2 - \bar{m}^2}. $$
The term in square brackets in Eq. (\ref{del_Z2_fin}) will multiply the 
Born cross section. The term proportional to $ ( \pm \not{p} - m )$ is 
identical to the like terms appearing in the expression for the 
vertex correction Eq. (\ref{vertex}). Since $  \Sigma_{ir}$ is the part 
of the self-energy correction which would be infrared divergent on-shell
(which means that it has a logarithmic resonant behavior
$$ \Sigma_{ir}(p^2) \sim \hbox{log}(p^2 - \bar{m}^2)$$
in the off-shell case), this term is also kept.

\subsection{Gauge invariance and corrections to particular subprocesses}

The partial amplitudes appearing in Eq. (\ref{mvg_amp}) can be directly related 
to Feynman diagrams and are straightforward to evaluate. 
However, as mentioned before, they cannot be directly identified with 
corrections to particular subprocesses. For example, the top - antitop
production vertex diagram (Figure \ref{vir_diag}a)) contributes to the
correction to production vertex, as well as to interference between
production and decay and even to interference between top decay and 
antitop decay, depending on when the top quark propagators are closer
to the resonances. Moreover, the amplitudes in Eq. (\ref{mvg_amp})
are not gauge invariant one by one, although their sum is.

For  purposes related to gauge invariance, and in order to be able to perform 
comparisons with the on-shell computation,
 it is desirable to 
decompose the total amplitude into gauge invariant corrections to 
particular subprocesses, and interference between these. 
The aim is to rewrite  Eq. (\ref{mvg_amp}) as:
\bec{m1a_gi}
{\cal{M}}^{vg} = {\cal{M}}_{prod} + {\cal{M}}_{tdec} + {\cal{M}}_{\bar{t}dec} +
{\cal{M}}_{prod-tdec}^{intf} + {\cal{M}}_{prod-\bar{t}dec}^{intf} + 
{\cal{M}}_{tdec-\bar{t}dec}^{intf}
\eec
with each term being gauge invariant by itself.

 To this end, it is necessary to decompose the amplitudes ${\cal{M}}_{tt}, {\cal{M}}_{tb} ...$
into parts which contribute solely to corrections to production, decay, or
interference. 
This decomposition will be based on the top and antitop
propagator structure of the matrix element. 
Following the prescription used in \cite{hardglu},
products of propagators
which go on-shell in different regions of the phase space can be decomposed 
as follows:
\bec{gi_dec}
\frac{1}{D(p_t)}\ \frac{1}{D(p_t-k)} \ = \
\frac{1}{D_0(p_t-k)}\ \frac{1}{D(p_t)}\ - 
\frac{1}{D_0(p_t-k)}\ \frac{1}{D(p_t-k)}
\eec
with 
\bec{prop_def}
 D(p) = p^2 - \bar{m}^2 \ , \ D_0(p-k) = (p-k)^2 - p^2
\eec
In Eq. (\ref{gi_dec}) the first term on the right hand side
is considered as a contribution to the production 
process and the second one a contribution to the decay process.
Furthermore, it is convenient to write the result in term of products
of gauge invariant currents (in a manner similar to \cite{ddr}).
For example, the ${\cal{M}}_{b\bar{t}}$ amplitude can be written
(using the extended soft gluon approximation):
\bec{gfgds}
{\cal{M}}_{b\bar{t}}(ESGA) = \frac{\alpha_s}{4\pi}\ {\cal{M}}_0 
\int \frac{d^4k}{i\pi^2}\ G_{\mu \nu}(k)\
\frac{ -2p_{\bar{t}}^{\mu} } { D(p_{\bar{t}}+k) }\
\frac{2p_b^{\nu}}{D_0(p_b-k)}\ \frac{D(p_t)}{D(p_t-k)}
\eec
where $G_{\mu \nu}(k)$ is the gluon propagator in an arbitrary gauge.
By using the currents:
\bec{gkfd}
j_{tdec}^{\ b ,\ \mu} = \frac{2p_b^{\mu}}{D_0(p_b-k)}\ \frac{D(p_t)}{D(p_t-k)}
\eec
$$
j_{prod}^{\ \bar{t},\ \mu} = \frac{-2p_{\bar{t}}^{\mu}}{D_0(p_{\bar{t}}+k)}\
\hbox{~~~,~~~}
j_{\bar{t}dec}^{\ \bar{t},\ \mu} = 
\frac{ -2p_{\bar{t}}^{\mu} }{ D_0(p_{\bar{t}}+k) }\
\frac{ D(p_{\bar{t}}) }{ D(p_{\bar{t}}+k) }
$$
the expression (\ref{gfgds}) can be written as:
\bec{mbbart_intf}
{\cal{M}}_{b\bar{t}}(ESGA) = \frac{\alpha_s}{4\pi}\ {\cal{M}}_0 
\int \frac{d^4k}{i\pi^2}\ G_{\mu \nu}(k)\
(\ j_{prod}^{\ \bar{t},\ \mu} - j_{\bar{t}dec}^{\ \bar{t},\ \mu} \ )
j_{tdec}^{\ b ,\ \nu} 
\eec
where the first term in parentheses contributes to production-decay
interference, and the second one contributes to decay-decay interference.

We can similarly define the currents:
\bec{glkvd}
j_{\bar{t}dec}^{\ \bar{b} ,\ \mu} = 
\frac{-2p_{\bar{b}}^{\mu}}{D_0(p_{\bar{b}}+k)}\ 
\frac{D(p_{\bar{t}})}{D(p_{\bar{t}}+k)}
\eec
$$
j_{prod}^{\ t,\ \mu} = \frac{2p_t^{\mu}}{D_0(p_t-k)}\
\hbox{~~~,~~~}
j_{tdec}^{\ t,\ \mu} = \frac{2p_t^{\mu}}{D_0(p_t-k)}\
\frac{ D(p_t) }{ D(p_t-k) }
$$
and the amplitudes for the other two interference diagrams can be written like:
\bec{mtbarb_intf}
{\cal{M}}_{t\bar{b}}(ESGA) = \frac{\alpha_s}{4\pi}\ {\cal{M}}_0 
\int \frac{d^4k}{i\pi^2}\ G_{\mu \nu}(k)\
(\ j_{prod}^{\ t,\ \mu} - j_{tdec}^{\ t,\ \mu} \ )
j_{\bar{t}dec}^{\ \bar{b} ,\ \nu}
\eec
\bec{mbbarb_intf}
{\cal{M}}_{b\bar{b}}(ESGA) = \frac{\alpha_s}{4\pi}\ {\cal{M}}_0 
\int \frac{d^4k}{i\pi^2}\ G_{\mu \nu}(k)\
j_{tdec}^{\ b ,\ \mu} \ j_{\bar{t}dec}^{\ \bar{b} ,\ \nu} \ \ .
\eec

Contributions to interference between subprocesses do not come solely
from the manifestly non-factorizable diagrams. The diagrams in which
the gluon contributes to vertex or self-energy corrections (Fig. 3a))
also contain interference terms. 
Since the decomposition into purely vertex (or self-energy) corrections
and interference corrections is not unique, we shall 
present our approach in some detail.

The amplitude for the vertex correction diagram with off-shell particles
can be written as:
\bec{kaglkad}
 (\delta \Gamma)_{12} = \frac{\alpha_s}{4 \pi} \
\int \frac{d^4 k}{i \pi^2}\ G_{\mu \nu}(k)\ \gamma^{\nu}\
\frac{A(p_1,p_2) + k^{\alpha}B_{\alpha}(p_1,p_2) + 
k^{\alpha}k^{\beta} C_{\alpha \beta}(p_1,p_2)}
{( (p_1 - k)^2 - m_1^2 )\ (p_2 + k)^2 - m_2^2 )}\
\gamma_{\nu}
\eec
A corresponding on-shell approximation for this amplitude would be 
\bec{kfgg}
 (\delta \Gamma)_{os} = \frac{\alpha_s}{4 \pi} \
\int \frac{d^4 k}{i \pi^2}\ G_{\mu \nu}(k)\ \gamma^{\nu}\
\frac{A(p'_1,p'_2) + k^{\alpha}B_{\alpha}(p'_1,p'_2) + 
k^{\alpha}k^{\beta} C_{\alpha \beta}(p'_1,p'_2)}
{( -2p'_1 k + k^2  )\ ( 2 p'_2 k + k^2 )}\
\gamma_{\nu}
\eec
where $p'_1$ and $p'_2$ are some on-shell approximations for 
$p_1$ and $p_2$ ($p_1^2 = p_2^2 = m^2$).
 Now, we can define the interference contribution through:
\bec{jdhgdso}
(\delta \Gamma)_{12} = (\delta \Gamma)_{12}^{os} + 
(\delta \Gamma)_{12}^{intf}
\eec
Note, however, that $(\delta \Gamma)_{12}^{os}$ is not unique, 
since $p'_1,p'_2$
are not unique; different choices for these momenta would yield different
results for $(\delta \Gamma)_{12}^{os}$.
 The uncertainty which arises is, of course, of order
$p^2-m^2$, so it can be neglected in the DPA. However, it allows us to 
choose the following definition for $(\delta \Gamma)_{12}^{os}$:
$$
(\delta \Gamma)_{12}^{os} = \frac{\alpha_s}{4 \pi} 
\int \frac{d^4 k}{i \pi^2}\ G_{\mu \nu}(k) \left[
\frac{(2p_1^{\mu}) \ {\Gamma}\ (-2p_2^{\nu})}{ D_0(p_1-k) D_0(p_2+k)}
\right.
$$
\bec{ga_os} \left. +\
\gamma^{\mu} \frac{ k^{\alpha}B_{\alpha}(p_1,p_2) + 
k^{\alpha}k^{\beta} C_{\alpha \beta}(p_1,p_2)}
{ D(p_1 - k) D(p_2 + k) } \gamma^{\nu} \right] .
\eec
This choice means that the purely vertex correction (factorizable) part of the 
vertex diagram can be obtained by simply replacing the off-shell 
$C_0(p_1,p_2,0,m_1,m_2)$
function appearing in the expression for $(\delta \Gamma)_{12}$
with the on-shell, infrared divergent function 
$C_0(p_1,p_2,\mu,\sqrt{p_1^2},\sqrt{p_2^2})$.

Conversely, the interference part of the off-shell vertex correction diagram
is:
\bec{kohdsi}
(\delta \Gamma)_{12}^{intf} = 
\frac{\alpha_s}{4 \pi} 
\int \frac{d^4 k}{i \pi^2}\ G_{\mu \nu}(k) \left[
\gamma^{\mu} \frac{ A(p_1,p_2) }{ D(p_1 - k) D(p_2 + k) }
\gamma^{\nu} - 
\frac{(2p_1^{\mu}) \ {\Gamma}\ (-2p_2^{\nu})}{ D_0(p_1 - k) D_0(p_2 + k)}\right]
\eec
with $ A(p_1,p_2) = (\not{p_1} + m_1)  {\Gamma} (-\not{p_2} + m_2)$.
For the $t \bar{t}$ production diagram, \nolinebreak[5] in \nolinebreak[5]
 DPA 
$$ \gamma^{\mu} (\not{p_t} + \bar{m}_t) \ {\Gamma_{t \at}}\ 
(-\not{p}_{\bar{t}} + \bar{m}_t) \gamma^{\nu} \rightarrow
(2p_t^{\mu}) \ {\Gamma_{t \at}}\ (-2p_{\bar{t}}^{\nu})
$$
leading to 
{\small \bec{lgdvkj}
{\cal{M}}_{t\bar{t}}^{intf} = \frac{\alpha_s}{4\pi}\ {\cal{M}}_0 
\int \frac{d^4k}{i\pi^2}\ G_{\mu \nu}(k)\ \left[
(- j_{tdec}^{\ t,\ \mu} )\ j_{prod}^{\ \bar{t},\ \nu}\ +
j_{prod}^{\ t,\ \mu} (- j_{\bar{t}dec}^{\ \bar{t},\ \mu} )\ +
(- j_{tdec}^{\ t,\ \mu} )\ (- j_{\bar{t}dec}^{\ \bar{t},\ \mu} )
\right] .
\eec }

Things are  different for the decay vertex corrections, since
we have doubly resonant contributions which are not proportional
to the tree level matrix element. In the 
top decay case, the transformation:
$$ \gamma^{\mu} (\not{p_b} + \bar{m}_b) \ \not{\epsilon}_{W^+}\ 
 (\not{p_t} + \bar{m}_t) \gamma^{\nu} \rightarrow
(2p_b^{\mu}) \  \not{\epsilon}_{W^+}\  \left[ 2p_t^{\nu}
+ \gamma^{\nu} (-\not{p_t} + \bar{m}_t) \right]
$$
will lead to:
\bec{vhgdsj}
{\cal{M}}_{t b}^{intf} = \frac{\alpha_s}{4\pi}\
\int \frac{d^4k}{i\pi^2}\ G_{\mu \nu}(k)\ 
j_{tdec}^{\ b ,\ \mu}\ \left[ (-j_{prod}^{\ t,\ \nu}) {\cal{M}}_0 +
M_1^{t, \ \nu} \right]
\eec
where
\bec{gkfdsah}
M_1^{t, \ \mu}\ = \ 
\frac{-1}{D(p_t)\ D(p_{\bar{t}})}\ \left[
\bar{u}(b) \not{\epsilon}_{W^+}\
\gamma^{\mu}\ {\Gamma_{\gamma,Z_0}}\  (-\not{p}_{\bar{t}} + \bar{m}_t)
\not{\epsilon}_{W^-} v(\bar{b}) \right]
\eec
In a similar manner, the interference term ${\cal{M}}_{\bar{t} \bar{b}}^{intf}$ 
coming from the antitop vertex correction diagram can be written in terms 
of the currents $j_{\bar{t}dec}^{\ \bar{b} ,\ \mu},\ 
-j_{prod}^{\ \bar{t},\ \nu}$, and the matrix element $M_1^{\bar{t}, \ \nu}$.

Finally, the last diagrams to be split into on-shell and interference 
contribution are the top, antitop self-energy diagrams. Using the same 
approach as in the vertex case, we define:
$$ 
(\Delta Z)_{t}^{intf} = \frac{\alpha_s}{4\pi}\  \left\{
\int \frac{d^4k}{i\pi^2}\ G_{\mu \nu}(k)\ \left[
\gamma^{\mu} \frac{\not{p_t} + \bar{m}_t}{D(p_t-k)} \gamma^{\nu}\
-\ \frac{2p_t^{\mu} \gamma^{\nu}}{ D_0(p_t-k)} \right] \right\}\ 
\frac{\not{p_t} + \bar{m}_t}{D(p_t)}\ 
$$
\bec{jhvdsj}
-\ (-1)\frac{\alpha_s}{4\pi}\ 
\int \frac{d^4k}{i\pi^2}\ G_{\mu \nu}(k)\
\frac{2p_t^{\mu}}{D_0(p_t-k)} \frac{2p_t^{\nu}}{D_0(p_t-k)}
\eec
where the quantity in the the curly brackets is the renormalized top
self-energy, and the quantity on the second line is the on-shell limit
of the quantity on the first line. 
This will lead to the following result for the interference 
contribution coming from the top self-energy diagram:
\bec{lkjgds}
{\cal{M}}_{tt}^{intf} = \frac{\alpha_s}{4\pi}\ 
\int \frac{d^4k}{i\pi^2}\ G_{\mu \nu}(k)\
\left[ j_{prod}^{\ t,\ \mu}\ {\cal{M}}_0\ -\ M_1^{t, \ \mu} \right]
j_{tdec}^{\ t,\ \nu}
\eec
and a similar one from the antitop self-energy diagram.

Now we have all the pieces needed to write down the interference terms.
The final result is:
$$
{\cal{M}}^{intf} = \frac{\alpha_s}{4\pi}\ 
\int \frac{d^4k}{i\pi^2}\ G_{\mu \nu}(k)\ \left[
(j_{prod}^{\ \mu} {\cal{M}}_0 + M_1^{t, \ \mu}) j_{tdec}^{\ \nu} - 
(j_{prod}^{\ \mu} {\cal{M}}_0 - M_1^{\bar{t}, \ \mu}) j_{\bar{t}dec}^{\ \nu}
\right.
$$
\bec{gi_intf_dec}
\left.
+ j_{tdec}^{\ \mu} j_{\bar{t}dec}^{\ \nu} {\cal{M}}_0 \right]
\eec
It is easy in this formula to identify the production-decay or decay-decay
interference terms. The currents:
\bec{dkscdsa} j_{prod} = j_{prod}^{\bar{t}} - j_{prod}^{t} \mbox{~~,~~}
j_{tdec} = j_{tdec}^b - j_{tdec}^t \mbox{~~,~~}
j_{\bar{t}dec} = j_{\bar{t}dec}^{\bar{b}} - j_{\bar{t}dec}^{\bar{t}}
\eec
are conserved, and gauge invariant in DPA.
 Therefore, the total interference contribution
as well as the interference between subprocesses parts are gauge invariant
in the approximation used.

\section{Computational Approach}

Once a consistent scheme for evaluating the virtual corrections 
to the top production and decay process (\ref{tree_p}) has been set up
(as described in the previous sections),
the next step is the implementation of this scheme
in a Monte Carlo program.  
In this section we give some details about the technical issues 
arising in the design of such a program, and how we choose to solve them.

There are two types of quantities involved in the evaluation of the
NLO amplitude: scalar quantities (form factors), which encode 
the contribution of loops, and spinorial quantities,
built from Dirac spinors and operators.
For example,
the contribution coming from the $t \bar{t}$ vertex correction can be written:
\bec{cdsa}
\tilde{\cal{M}}_{t \bar{t}} = 
\bar{u}(p_b) \not{\epsilon}_{W^+}  (\not{p}_t + m_t) \
\delta \Gamma^{t\at}_{ren} (-\not{p}_{\bar{t}} + m_t) 
\not{\epsilon}_{W^-}  v(p_{\bar{b}})
\eec
or, using the decomposition in Eq. (\ref{vertex}) :
\bec{prod_cont}
\tilde{\cal{M}}_{t \bar{t}}
 = \frac{\alpha_s}{4 \pi} \sum_{i = 1,2} \left[
C_V\ F_i^V\ T_i^V\ + C_A\ F_i^A\ T_i^A \right] 
\eec
(for more details see \cite{thesis}).

Let's start by discussing the evaluation of the scalar form factors $F_i$.
Rather than compute analytic expressions for each form factor, we have
chosen to evaluate them in terms of Passarino-Veltman (PV)
functions \cite{passa}. This approach has the
advantage that we have to compute only a
 few quantities which contain logarithms 
and dilogarithms: the ${\cal{B}}_0$ two-point and ${\cal{C}}_0$ three-point 
scalar functions (all the rest can be
written as linear combinations of these functions). In turn, for evaluating
the  PV 2 and 3-point scalar functions, we use the FF routines \cite{old}. 

To compute the amplitudes corresponding to the interference diagrams, we 
need to be able to evaluate the 4-point and 5-point scalar integrals in 
Eqs. (\ref{mbbtr}), (\ref{mtbbr}), (\ref{mbbbr}).
 There are no published results (or routines) for the
general (complex masses) 4-point scalar integrals. We have build such
routines for the infrared finite ${\cal{D}}_0$
 function by using the general methods
described in \cite{thooft_si}. The results of these routines have been checked 
against analytical results in the soft gluon approximation published in 
\cite{BBC}. 

The 5-point scalar function ${\cal{E}}_0$
 has been computed by reduction to 4-point
functions, following the recipe in \cite{ddr}. The resulting infrared divergent
4-point functions have been evaluated using the analytic results published
in \cite{denner_ir4pt}.

Some comments on the treatment of the top width are needed here. One way of 
evaluating the scalar form factors in Eq. (\ref{prod_cont}) is to compute the
gluon integrals in the zero top width limit and introduce the finite width only 
in terms which are divergent on-shell (that is, replace $m_t^2$ with 
$\bar{m}_t^2 = m_t^2 - i m_t \Gamma_t$ in terms like log$(p_t^2 - m_t^2)$;
see for example \cite{ddr}).
The difference between this result and the one obtained by using
the complex top mass in all the terms is of order $\Gamma_t/m_t$, therefore
at about 1\% level. This would be acceptable if the radiative corrections 
would be small with respect with the tree level result (as is the case for the
$W$ production process), but in our case it turns out that the one-loop
QCD corrections are of the same order of magnitude as the tree level 
result\footnote{The reason the {\it total } QCD corrections are of 
order 10 - 20\%
is because of large cancellations between the virtual corrections and
soft gluon real corrections.}. Therefore, order \% terms are important. Since
in the case of real gluon radiation the top width appears in all terms, for 
reasons of consistency we need to keep the width 
in all terms in the evaluation of the virtual corrections too. 

 The other elements needed in the evaluation of the amplitude (\ref{prod_cont})
are the spinor sandwiches $T_i$. We compute these quantities using
spinor techniques, as for the real gluon radiation case. Since 
this part of the computation is quite complex, and hence prone to errors, we 
have two different ways of performing it. In one approach, we express the
$T_i$'s in terms of basic spinor products
$ \bar{u}(p_i,s_i) u(p_j,s_j)$; this is the more involved
case (in terms of the work done by the programmer), geared  for
implementation in a Fortran routine, and which allows fast computation. 
The other approach uses C++ routines which allow the 
automated evaluation of general spinor sandwiches like 
$$ \bar{u}(p,s)(\not{p}_1 + m_1)(\not{p}_2 +m_2) \ldots u(p',s')$$
(To this purpose, we have 
constructed classes that describe $<bra|$ and $|ket>$ spinors,
and operators of type $ \not{p}_i \pm m_i $; in turn, 
these classes use the basic classes - 4-vector, complex number - 
defined in the 
Pandora event generator \cite{peskin}).
This method allows easy evaluation of $T_i$ expressions
(again from the programmer's viewpoint)
and is much more resistant to programming errors;
 but the computation is slower than in the previous method. Therefore, 
the main use of the results obtained from  the C++ routines is 
to check the Fortran results.

\section{On-shell DPA}

The issue of interference effects in the production and decay 
of heavy unstable particles has been the subject of extensive
studies in the past decade. One of the main results
is the theorem, due to Fadin, Khoze and Martin \cite{FKM_0}, which states
that these interference effects are suppressed
(see also a more recent discussion in \cite{CKSS}).  A stronger version
of this theorem \cite{FKM_1} claims that NLO interference effects
cancel in inclusive quantities up to terms of order $\alpha \Gamma/M$.
As in \cite{FKM_1}, it is possible to 
define and use a framework
for the computation of interference corrections in which the total 
interference contribution to inclusive quantities is zero.


In order to facilitate a comparision with our results (where the interference
contributions do not cancel completely), 
we shall discuss this alternative approach 
(which we shall call on-shell DPA) in some detail.
Results obtained using on-shell DPA have been obtained for 
the $W$ pair production case
at LEP II (\cite{ddr}, \cite{BBC}), as well as, more recently,
for the evaluation of interference (non-factorizable) 
corrections to the top production and decay process at $e^+ e^-$ 
and hadron colliders \cite{bbc_top}.

The relevant features of this approach are two: first, the amplitudes
for corrections to subprocesses are computed in the on-shell approximation
for the top quarks.
For example, the correction to the production process can be written in 
terms of the on-shell amplitude:
\bec{os_prod}
\tilde{\cal{M}}_{prod}^{os} \ = \ \sum_{\la,\la'}
{\cal{M}}_{\la,\la'}(e^+e^- \to t \at (g))\
{\cal{M}}_{\la}(t \to b W^+)\ {\cal{M}}_{\la'}(\at \to \ab W^-)
\eec
where $\la,\la'$ are the spins of the top quarks.
The difference between the above amplitudes and the ones used in this 
paper (Eqs. (\ref{m1a_gi}), (\ref{ga_os})) is due to non-doubly 
resonant terms, therefore acceptable in DPA.

The other characteristic feature of the on-shell DPA method
is that the interference due to real gluon radiation is computed
by using a semianalytic approach. This approach rests on
the observation that interference is due mainly to gluons of energies
of order $\Gamma_t$; therefore, the (extended)
soft gluon approximation can be employed for
the evaluation of interference terms.

There are two stages where this approximation comes into play. First,
it is used at the matrix element evaluation level. For example,
consider interference between the diagrams where the gluon is radiated
from the bottom quark and from the antitop quark:
\bec{tbb_intf}
d\sigma^{rg}_{\at b}(p_b,p_W,\ldots,k) \sim 
2 \hbox{Re}\left[
\sum_{\epsilon_g}\ {\cal{M}}_{b}^{sg}\ ({\cal{M}}_{\at}^{sg})^* \right]
\eec
$$
\ = \ |{\cal{M}}_0(p_b,p_W,\ldots)|^2 \
2 \hbox{Re} \left[ \ \frac{4p_{\at} p_{b}}{2kp_{b}+i\epsilon}\
\frac{p_{t}^2 - \bar{m}_t^2}{(p_{t}+k)^2 - \bar{m}_t^2}\
\frac{1}{(p_{\at}+k)^2 - \bar{m}_t^{2*}} \ \right].
$$

 The second stage is the treatment of the final state phase space. In the 
soft gluon approximation, it factorizes: 
$d\Omega_{b,W,\ldots,g} = d\Omega_{b,W,\ldots} \times d\Omega_g$ , and
the integration over the gluon momenta is performed separately:
$$
d\sigma^{rg}_{\at b}(p_b',p_W',\ldots)\ = \  
|{\cal{M}}_0(p_b',p_W',\ldots)|^2 \ 
$$
\bec{tbb_sig} \times \
\frac{\alpha_s}{\pi} \hbox{Re} \left[ \int \ \frac{d^3k}{2\pi \omega}
\ \frac{4p_{\at} p_{b}}{2kp_{b}+i\epsilon}\
\frac{p_{t}^2 - \bar{m}_t^2}{(p_{t}+k)^2 - \bar{m}_t^2}\
\frac{1}{(p_{\at}+k)^2 - \bar{m}_t^{2*}} \ \right]
\eec
where $p_b',p_W',\ldots$ are given by a suitable projection of the
off-shell momenta  $ p_b,p_W,\ldots $ onto the on-shell phase space (for
an example of how this projection might be accomplished see \cite{bbc_details}).
In the above equation  $p_t = p_b' + p_{W^+}', \ 
p_{\at} = p_{\ab}' + p_{W^-}'$, and the integral over gluon momenta
is allowed to go to infinity (since hard gluons contribute nonresonant
terms to the result).

The quantity on the second line of Eq. (\ref{tbb_sig}) can be evaluated 
analytically, through methods similar to those used to evaluate the
virtual 4-point functions. We will not give the results here 
(they can be found in \cite{ddr}, \cite{BBC}),
but there is an important comment to
make. Using this procedure to compute the real gluon interference,
the total interference obtained by adding the virtual diagram contribution
(Eq. (\ref{mbbtr})) to the above result and integrating over the top invariant
mass parameter is zero \cite{FKM_1}.
This cancellation works also for the other interference diagrams; therefore,
in this approach, the contribution of non-factorizable corrections is 
zero to the total cross section. However, this result depends on two 
things. First, it requires an inclusive treatment of real gluon radiation, with 
phase space integration extending to infinity. Second, both the virtual
and the real interference terms have to be treated in the soft gluon
approximation.

But, is the use of the soft gluon approximation justified in this case?
At the amplitude level (Eq. (\ref{tbb_intf})), the answer is yes; the relevant 
gluon energy, being of order $\Gamma_t$, is much smaller than the other
momenta involved. However, this approximation does not seem to be acceptable
for the phase space factorization stage of the above approach. Here, problems
might arise when projecting 
the off-shell momenta onto the on-shell phase space. The reason is that
there is no single way to perform this projection; therefore, in the 
determination of the on-shell momenta $p_b', p_W', \ldots$ there is an 
uncertainty of the order of the gluon energy, or $\Gamma_t$. Now, 
being close to the top resonances,
we are in a region of the phase space where the cross section varies
greatly over a range of energy of order $\Gamma_t$
(due to the top quark propagators); therefore such an 
uncertainty is not acceptable.

To illustrate the dependence of the result for real gluon interference
on the choice of the on-shell momenta $ p_b', p_W', \ldots$, let's presume
that instead of projecting $p_b$ into $p_b'$, we also take into account
the gluon momentum: $p_b + k \to p_b'$ (physically, this might be justified by 
the inclusion of the gluon jet in the bottom quark jet). 
Then, Eq. (\ref{tbb_sig}) becomes:
\bec{tbb_sig1}
d\sigma^{rg}_{\at b}(p_b',p_W',\ldots)\ = \  
|{\cal{M}}_0(p_b',p_W',\ldots)|^2 \ 
\frac{\alpha_s}{\pi} \hbox{Re} \left[ \int \frac{d^3k}{2\pi \omega}
\ \frac{4p_{\at} p_{b}}{2kp_{b}+i\epsilon}\
\frac{1}{(p_{\at}+k)^2 - \bar{m}_t^{2*}} \right]
\eec
The result for the above expression is different from the result 
for Eq. (\ref{tbb_sig}), and the difference contains doubly resonant terms.
Therefore, in the on-shell DPA approach, the result for the interference
terms depends on the choice of the implementation of the
 phase space factorization. A discussion
of this dependence for the $W$ pair production case can be found in \cite{ddr}.

\section{Results for virtual corrections and the total cross section}
 
In this section, we present some results on the total cross section
for the top production and decay process at linear colliders. We take into 
account the virtual corrections as well as contributions coming from 
real gluon radiation. Furthermore, we study the effect of 
interference (nonfactorizable)
terms on invariant top mass distributions and  we perform comparisons
with results previously published \nolinebreak[3] \cite{bbc_top}.

In obtaining the results presented in this section, 
the following set of parameters is used:
$$ m_t = 175\ \mbox{GeV,} \mbox{~~~~}
\alpha_s = 0.1, \mbox{~~~~}
\Gamma_{t}^0 = 1.55\ \mbox{GeV,} \mbox{~~~~}
\Gamma_{t} = 1.42\ \mbox{GeV,} \mbox{~~~~}
$$
where $\Gamma_{t}^0$ is the top width at the tree level, while
$\Gamma_{t}$ includes the NLO QCD radiative corrections.

\vspace{0.3cm}
We start by looking at the total cross section for our process. 
Table 1 presents results for the following quantities:
\begin{itemize}
\item 
$\sigma_0$ : cross section for the tree level process (\ref{proc1});
computed in the on-shell (narrow width) approximation,
using the zero-order top width $\Gamma_{t}^0$.
\item 
$\sigma_1^{os}$ : cross section for the NLO process in the on-shell 
approximation; computed using the NLO top width $\Gamma_{t}$. 
\item 
$\sigma_1^{fact}$ : the main (factorizable) part of the DPA approximation
to the NLO process.
This quantity contains corrections to production and decay as defined
in section 2.4. 

\item 
$\sigma_1^{intf}$ : the interference (non-factorizable) part of the DPA
approximation to the NLO process, as defined in section 2.4.
\end{itemize} 

\begin{table}[!h]
\begin{center}
\begin{tabular}{|c|r @{.} l|r @{.} l| r @{.} l c|}
\hline
 	& \multicolumn{2}{c}{360 GeV} &  \multicolumn{2}{c}{500 GeV} & 
 	\multicolumn{2}{c}{1000 GeV} &\\
\hline
 $\sigma_0 (pb)$       & 0 & 386& 0 & 570 & 0&172 &\\
 $\sigma_1^{os}(pb)$   & 0 & 700(1)& 0 & 660(2) & 0&1839(7) &\\
 $\sigma_1^{fact}(pb)$ & 0 & 676(2)& 0 & 664(2)  & 0&1920(8) &\\
 $\sigma_1^{intf}(pb)$ & -0 & 032(1)& -0 & 0116(3) & -0&0061(2) &\\
\hline
\end{tabular}
\end{center}
\caption{Total cross sections for top production at linear colliders 
with no cuts on phase space (numbers in parantheses are errors due to 
numerical integration).}
\label{table1}
\end{table}

We present results for three values of collision center-of-mass energies:
360 GeV, just above the $t \at$ production threshold, 500 GeV, the most 
common value used in linear collider studies, and 1 TeV, which is relevant
for higher energy machines. The 360 GeV  result
is probably not good; this close to the threshold, resummation
of large logarithms appearing in the $t \at$ interaction 
is needed \cite{top_thres}. However, 
it is interesting to see the magnitude of the nonfactorizable corrections 
at fixed order in this energy range.

The NLO cross sections contain contributions from the virtual corrections as
well as from real gluon radiation (evaluated as in \cite{hardglu}).
The phase space splicing method is 
used for the treatment of infrared singularities.
The value of the technical cut which separates the infrared from the real 
gluons is $\epsilon = 0.1$ GeV;
the results are independent
of the choice of this parameter\footnote{Taking its value much 
smaller than the top width allows the use of the 
soft gluon approximation and phase space factorization 
in the evaluation of the infrared integrals.}.
No physical cuts have been imposed on the final phase space.

The results in Table 1 prompt several comments.
First, the difference between $\sigma_1^{os}$ and $\sigma_1^{fact}$ 
is due to non-doubly-resonant terms, therefore it could be expected 
to be small. This is indeed the case at 500 GeV; but at 1 TeV, this difference
is about 4\% of the cross section. The reason is that in obtaining these
 results, we have integrated over the complete kinematic range available for
the top quark invariant mass (that is, 
$ m_b + m_W < \sqrt{p_t^2} < W - (m_b + m_W)$), so 
the cross-sections include contributions 
from regions of the phase space where the top quarks 
are far off-shell and non-resonant
terms are important. 

\begin{table}[!ht]
\begin{center}
\begin{tabular}{|c|r @{.} l|r @{.} l| r @{.} l c|}
\hline
 	& \multicolumn{2}{c}{360 GeV} &  \multicolumn{2}{c}{500 GeV} & 
 	\multicolumn{2}{c}{1000 GeV} &\\
\hline
 $\sigma_1^{os}(pb)$   & 0 & 682(1)& 0 & 627(2) & 0&1742(7) &\\
 $\sigma_1^{fact}(pb)$ & 0 & 670(2)& 0 & 629(2)  & 0&1781(5) &\\
 $\sigma_1^{intf}(pb)$ & -0 & 034(1)& -0 & 0068(2) & -0&0017(1) &\\
\hline
\end{tabular}
\end{center}
\label{table2}
\caption{Total cross sections for top production at linear colliders
with cuts on the top, antitop invariant mass.}
\end{table}

 In Table 2 we present the cross section results obtained 
with a cut on the $t, \bar{t}$ invariant mass 
$ | \sqrt{p_t^2},\sqrt{p_{\bar{t}}^2} - m_t | < 15$ GeV. The difference
between the two results for the main terms $\sigma_1^{os}$ and
$\sigma_1^{fact}$ is smaller in this case. Note 
that, since in the on-shell approach $p_t^2, p_{\bar{t}}^2 = m_t^2$,
 $\sigma_1^{os}$ in Table 1 and 2
contains a factor which simulates 
the effect of cuts (either from kinematic constraints or imposed ones)
on the top invariant mass. Note also that these cuts are not imposed
{\it ad hoc}, but they arise rather naturally in the process of defining 
a $t, \bar{t}$ production event; it makes sense to require that the reconstructed mass of the $b, W$ pairs is close to the top mass in the definition of such an event. In this context, it is also worth noting that
the contribution coming from the phase space region where
either the $t$ or $\bar{t}$ is far off-shell (more that ten times 
the width) is quite sizable (around 5\% of the total cross section 
for center-of-mass energies greater than 500 GeV).

\vspace{0.3cm}
Another quantity of interest is the differential interference cross section
as a function of the top invariant mass. Even if the total interference 
contribution to the cross section is small (at about 1\% level),
 it can have larger effects in differential distributions since
it can be positive in certain regions of the phase space and negative 
in others. In 
particular, it can be important in the reconstruction of the top invariant
mass; since $d \sigma_1^{intf}$ tends to decrease as $\sqrt{p_t^2}$ increases,
the net effect
would be to shift the position of the Breit-Wigner peak to smaller invariant
mass values. This effect can be quantified by the following equation:
the shift in the mass is
\bec{mass_shift}
\Delta M_t \ =\ \left. \left( \frac{d \delta_{nf}}{dM_t} \right)
\right|_{M_t = m_t}
\frac{\Gamma_t^2}{8}
\eec  
where $M_t = \sqrt{p_t^2}$, and $\delta_{nf}$ is the ratio of the 
non-factorizable (interference) part of the cross section to the
Born cross section:
$$ \delta_{nf} \ =\ \frac{d\sigma_1^{intf}}{d\sigma_0}$$

\begin{figure}[!ht] 
\centerline{\epsfig{file=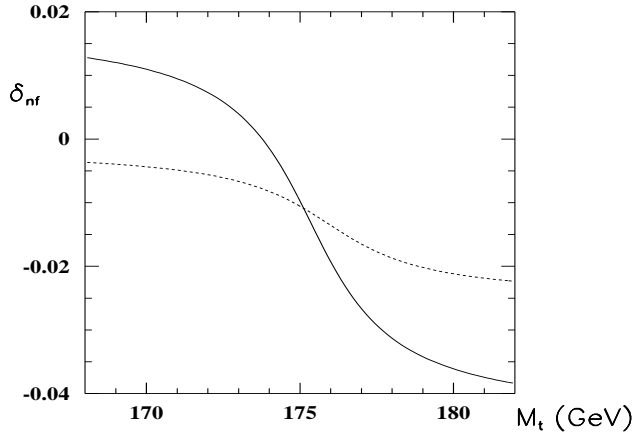,height=2.5in,width=3.in}}
\caption{ The relative nonfactorizable correction to the invariant
mass distribution; the solid line is the contribution of terms proportional
to the tree level amplitude, while the dashed line contains also the 
$M_1$ terms in Eq. (\ref{gi_intf_dec}). }
\label{interf_rat}
\end{figure}

In Figure \ref{interf_rat} we present the differential distribution
for the relative non-factorizable correction $\delta_{nf}(M_t)$ at center
of mass energy 500 GeV. The dashed line is the result which takes into 
account the full interference corrections in Eq. (\ref{gi_intf_dec}); the 
solid line is obtained by taking into account only the terms
proportional to the Born amplitude. Note that,
 although the contribution of the 
$M_1$ terms in Eq. (\ref{gi_intf_dec}) to the total cross section
is very close to zero, they have 
a sizable effect on the differential distribution in Fig. \ref{interf_rat}. 
Using Eq. (\ref{mass_shift}), we conclude that the shift 
in the position of the peak in the top invariant mass distribution 
due to interference effects is very small (of order of a few MeV -- 
in agreement with \cite{bbc_top}).

\begin{figure}[!b] 
\centerline{\epsfig{file=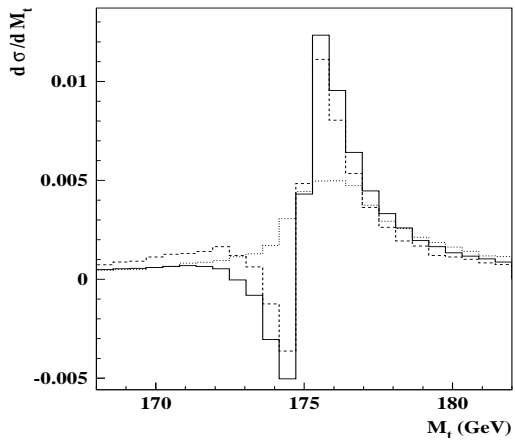,height=2.5in,width=3.in}}
\caption{ Real gluon interference: the $\sigma_{b\at}$ term. The
solid line corresponds to the semianalytic approach; the dashed line
is obtained through numerical evaluation with $M_t = \sqrt{p_{bW}^2}$;
the dotted line is obtained through numerical evaluation with $M_t$
given by Eq. (\ref{recomb}).}
\label{real_interf}
\end{figure}

The results in Table 1 and 2 indicate that
the contribution of interference terms to the total cross section is
of order 1\%, in agreement with the $\Gamma_t/m_t$ order
of magnitude  expected from naive arguments. However, it is not zero, as 
implied by results presented in \cite{bbc_top}, which use the on-shell
DPA method.
We have argued in section 4 that this difference is due to the way
in which the radiation of real gluon with energies of order $\Gamma_t$
is treated. In Figure \ref{real_interf},
we present the results for the real interference 
between the diagram where the gluon is radiated from the bottom quark
and the diagram where the gluon is radiated from the antitop quark.
The solid line is the result of the semianalytical approach described
in section 4. The other two lines are the result of the exact off-shell
computation (where the integration over the gluon momenta is performed
numerically). The two lines correspond to two different ways in which
the gluon momentum is treated in the reconstruction of the 
invariant top mass. For the dashed line, the gluon momentum is ignored
in the top mass reconstruction: $M_t = \sqrt{p_{bW}^2}$. Note that in this 
case, the result is quite close to that of the semianalytical computation, 
which is natural, since the gluon momentum is treated in both cases the same
way.

To obtain the dotted line, we have followed a more realistic approach, in 
which the gluon is included in the top mass reconstruction if
it happens to be radiated close enough to the top quark:
\bec{recomb} M_t \ = \ \left\{
\begin{array}{ll}
\sqrt{p_{bWg}^2} & \hbox{if }\ \hbox{cos}\theta_{tg}<\pi/3 \\
\sqrt{p_{bW}^2} & \hbox{otherwise}
\end{array} \right.
\eec
Although the total cross section is the same as for the other exact 
evaluation case, the differential cross section differs by quite a bit.

\begin{figure}[!b] 
\centerline{\epsfig{file=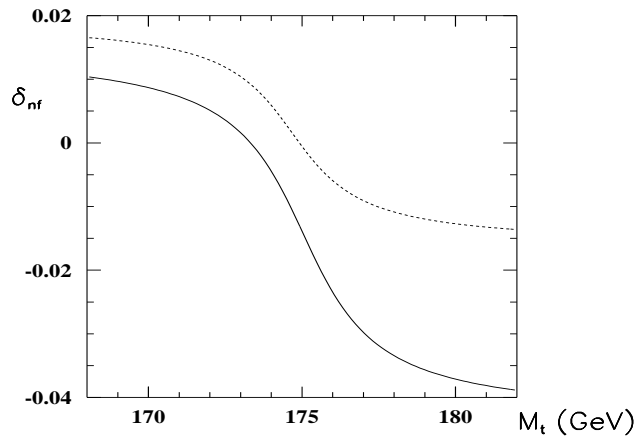,height=2.5in,width=3.in}}
\caption{ The relative nonfactorizable correction to the invariant
mass distribution; comparison between the semianalytical (dashed line) 
and the numerical (solid line) approach. }
\label{ir_comp}
\end{figure}

The total cross section corresponding to the 
 interference term presented in
Figure \ref{real_interf} has the value $\sigma_{b\at} = 0.121\ pb$ for the
semianalytical result, and $\sigma_{b\at} = 0.124\ pb$ for the numerical one.
Note that, since the diagram set contributing to this interference term is not
gauge invariant, this result has no physical meaning by itself. However, 
from these numbers we can get some insight  concerning the 
evaluation of interference corrections. First, note that the contribution
of this single diagram is much bigger (about two orders of magnitude) than
the total result for the interference terms. This means that there
are large cancellations taking place between the real and virtual interference 
contributions. This is quite natural, in accordance 
with the discussion in section 4; however, this also means that small
uncertainty (order percent) in evaluating one of this contributions
(the one coming from the real gluon interference, for example) can lead 
to large uncertainties in the evaluation of the total interference
contribution.

In Figure \ref{ir_comp} we present the comparison between the 
relative non-factorizable corrections computed in the semianalytical
approximation (dashed line) and the complete off-shell approach (solid line).
For the purpose of this comparison, we consider only the terms 
proportional to the Born amplitude in the exact computation, since
only these terms are taken into account in the semianalytical approach.
Note that the total interference cross section integrates to zero in the
latter case, and, as discussed above, the complete off-shell distribution
contains contributions that do not cancel in the total cross section. 


\section{Conclusions}

In this paper, we have discussed in some detail the evaluation of 
next to leading order QCD corrections to the top production and decay
process at a linear collider.
Since a full computation of the NLO amplitudes contributing to the 
process $e^+ e^- \rightarrow\ b\ W^+\ \bar{b}\ W^-$ is not feasible,
 the double pole approximation (DPA) has been employed. In our case this 
means taking into account only the diagrams which contain two intermediate top
quarks. Unlike most of the previous treatments, we allow for the two top
quarks to be off-shell, and include also the corrections due to 
interference between the top production and decay processes.

A parallel is drawn between our computation and the DPA evaluation 
of QED corrections to the $W$ pair production and decay process at LEP.
While there are many similarities between the two computations, 
there are also some differences; the most striking one is that
there are nonfactorizable (interference) corrections no longer proportional
to the Born amplitude in the top quark case.
Although the contribution of these terms to the total cross section (or other
inclusive quantities) is 
consistent with zero, they contribute to differential distributions. 

 Previous results obtained using a semianalytical approach 
state that the interference
terms cancel out completely in inclusive quantities 
(like the total cross section). We discuss the evaluation of the 
real gluon interference terms using analytic methods and we point out
the shortcomings of this approach.
The total magnitude of nonfactorizable corrections in our computation 
is found to be of order 1\% of the cross section.
We also present results for the total cross section
and the relative nonfactorizable correction to the top invariant 
mass distribution. The effect of nonfactorizable corrections on the 
top mass reconstruction is found to be very small.  

The results presented in this paper are relevant at collision energies
above the top -antitop production threshold.
Since we are interested in differential distributions of final state kinematic
variables, the approach used to obtain cross section distributions
is that of numerical
simulations. The amplitudes are evaluated using spinor techniques,
and the integration over the final state variables is performed using
Monte Carlo techniques. This approach has the added advantage
that it allows for the inclusion of experimentally relevant 
selection criteria on the final state phase space.

The calculation presented here is entirely at the parton level. For more
realistic simulations, it is necessary to take into account 
initial state related issues, like initial state radiation (ISR), beam
energy spread and beamstrahlung. The hadronization of the final 
state partons also has to be modeled.
Further extensions of interest include taking into account singly-resonant
diagrams (tree level as well as NLO in the on-shell approximation) and
the evaluation of electroweak corrections.



\vspace{0.7cm}
\noindent
{\Large \bf{Acknowledgments}}
\vspace{0.5cm}

I would like to thank Prof. L.H. Orr for support during the completion of this
work and for suggesting the topic in the first place. 
Also, many thanks to D. Wackeroth for discussions about the evaluation of
elecroweak corrections to the W pair production and decay process, to
A.P. Chapovsky for help with the numerical comparisions with previous results,
and to O.I. Yakovlev for clarifying some issues related to the DPA.  

This work was supported in part by the U.S. Department of Energy,
under grants DE-FG02-91ER40685 and DE-FG03-98ER-41076, 
and by the U.S. National Science Foundation, 
under grant PHY-9600155.

\vspace{0.7cm}
\noindent
{\Large \bf{Appendix} }

\appendix

\vspace{0.5cm}

In this appendix we shall give some details about  
the evaluation of vertex corrections and
fermion self energies. For a more extensive description, 
we refer the reader to \cite{thesis}. 

The amplitude for the general vertex correction in Figure \ref{ver_cor}
 can be written as: 
\bec{gen_gam}
 \delta \Gamma^{\mu} = \frac{\alpha_s}{4 \pi} \
\int \frac{d^4 k}{i \pi^2}\ \frac{1}{k^2}\ \gamma^{\nu}\
\frac{\not{p}_1 - \not{k} + m_1}{(p_1 - k)^2 - \bar{m}_1^2}\
\gamma^{\mu}(C_V + C_A \gamma^5 ) \
\frac{-\not{p}_2 - \not{k} + m_2}{(p_2 + k)^2 - \bar{m}_2^2}\
\gamma_{\nu}
\eec
Upon evaluation of the integral, the result can be written in terms of two
sets of form-factors; one for the vectorial part of the vertex correction, one
for the axial part. The number of form factors needed depends on the specific 
constraints on the process; in the general case, 
when the momenta $p_1, p_2$ are 
off-shell, eight$\times 2$ form factors are involved:
\bec{gen_gam_res}
\delta \Gamma^{\mu} = \frac{\alpha_s}{4 \pi}\
\sum _{i=1,8} \left( C_V F_i^V T_i^{V \mu} \ + \ C_A F_i^A T_i^{A \mu} \right)
\eec

The definition of these form factors depends on the choice of the spinorial
elements in terms of which the result is written. We
define:
\bec{del_gam_res}
\delta \Gamma^{\mu}_V = \frac{\alpha_s}{4 \pi} C_V\ \times
\eec
$$
\begin{array}{rclcccccl}
          [        & p_1^\mu & & F_1^V &  + 
 		&                   &  \gamma^{\mu} & & F_2^V \ + \\
(\not{p}_1 - m_1) & p_1^\mu & & F_3^V &  + 
		& (\not{p}_1 - m_1) &  \gamma^{\mu} & & F_4^V \ + \\
                  & p_1^\mu & (-\not{p}_2 - m_2) & F_5^V &  + 
 		& &  \gamma^{\mu} & (-\not{p}_2 - m_2) & F_6^V \ + \\
(\not{p}_1 - m_1) & p_1^\mu & (-\not{p}_2 - m_2) & F_7^V &  + 
& (\not{p}_1 - m_1) &  \gamma^{\mu} & (-\not{p}_2 - m_2)& F_8^V \ ]
\end{array}
$$
for the vectorial part of the vertex correction; for the axial one, replace
$p_1^\mu, \ \gamma^\mu$ with 
$p_1^\mu \gamma^5, \ \gamma^\mu \gamma^5$ in the expression above. This
definition has the advantage that when the particle $i$ is on-shell, the
terms which contain the $ \pm \not{p}_i + m_i$ drop out. Also, we have 
made use of the fact that, if $\delta \Gamma^{\mu}$ multiplies $A_{\mu}$ in 
the full matrix element ($A_{\mu}$ can be thought of as the polarization
vector of the gauge boson in diagram \ref{ver_cor}), then 
$(p_1+p_2)^{\mu} A_{\mu} = 0$.

 We shall evaluate and write the results for the form factors 
in terms of Passarino - Veltman functions
(for the definition of these see below):

\bec{f_vect}
F_1^V = 4[\ m_1(C_{12}-C_{11}+C_{23}-C_{21}) - m_2(C_{12}+C_{23}) \ ]
\eec
$$ F_2^V = -2[\ 2 p_1 p_2(C_0 + C_{11}) + 2(C_{24}-1/4) + 2B_0^{12} -1  
$$
$$
- m_1 m_2 C_{11}
+ p_1^2(-C_{11}+C_{12}) - p_2^2 C_{12} \ ] $$
$$ F_3^V = -4 ( C_0 + 2C_{11} - C_{12} + C_{21} - C_{23} ) \ \ , \ \ 
 F_4^V =  2 m_2 (C_0 + C_{11}) $$
$$ F_5^V = -4 ( C_0 + C_{11} + C_{23} + C_{12} ) \ \ , \ \ 
 F_6^V = 2 m_1 (C_0 + C_{11}) $$
 $$
F_7^V = 0 \ \ , \ \ F_8^V = 2 ( C_0 + C_{11} )
 $$

In the computation of the axial form factors,  $\gamma^5$ can be shifted
to the right in  Eq. (\ref{gen_gam}).
In the result for the form factors, this amounts to changing the sign of $m_2$,
$ m_2 \rightarrow -m_2$, and multiplying the $F_5, \ldots, F_8$
 with $(-1)$.
 Then :
\bec{f_axial}
 F_1^A = F_1^V + 8 m_2 (C_{12}+C_{23}) \ \ , \ \ 
 F_2^A = F_2^V - 4m_1 m_2 C_{11}
\eec
$$ F_3^A = \ \ F_3^V \ \ , \ \ 
 F_4^A = -F_4^V $$
$$ F_5^A = -F_5^V \ \ , \ \ 
 F_6^A = -F_6^V $$
$$ F_7^A = -F_7^V \ \ , \ \ F_8^A = -F_8^V$$

The result (\ref{del_Z2_fin}) for the fermion self-energy corrections
 can be similarly written in terms of Passarino-Veltman functions:
\bec{f_fse}
\Sigma_a(p^2) = \frac{\alpha_s}{4 \pi} (1 + 2B_1(p^2,\bar{m}^2))
\eec
$$
\Sigma_{ir}(p^2) = \frac{\alpha_s}{4 \pi} \frac{m^2}{p^2 - \bar{m}^2}
\left[ 4 \Delta B_0(p^2,\bar{m}^2) + 4 \Delta B_1(p^2,\bar{m}^2) \right]
$$
with 
$$ \Delta B_n(p^2,\bar{m}^2) = B_n(p^2,\bar{m}^2) - B_n(\bar{m}^2,\bar{m}^2)
\ , \ n = 0,1$$
If we further define the $X_0, X_1$ form factors through:
\bec{del_Z2_fin3}
\Delta Z_2(p) = 2\ \frac{\alpha_s}{4 \pi} \left[ X_0(p^2) \ + \ X_1(p^2)
( \pm \not{p} - m ) \right]
\eec
the renormalized vertex correction in Eq. (\ref{ren_ver2})
can be obtained by making the following redefinitions of form-factors in 
Eqs. (\ref{f_vect}), (\ref{f_axial}):
\bec{ren_f_fact}
F_2^{V,A} \ \to \ F_2^{V,A} + X_0(p_1^2) + X_0(p_2^2)
\eec
$$ F_4^{V,A} \ \to \ F_4^{V,A} + X_1(p_1^2)$$
$$ F_6^{V,A} \ \to \ F_6^{V,A} + X_1(p_2^2)$$

These general results are easily translated for the specific cases which
appear in our computation.  For corections to the production,
top decay and antitop decay vertices we have, respectively:

\bea
 F_{i,t \at}^{V,A} & \ = &\ F_i^{V,A}(
 p_1 = p_t, \ p_2 = p_{\bar{t}}, \ m_1 = m_2 = m_t )  \\
 F_{i,t b}^{V,A} & \ = &\ F_i^{V,A}(
p_1 = p_b, \ p_2 = -p_t ,\ m_1 = m_b, \ m_2 = m_t ) \nonumber\\
 F_{i,\at \ab}^{V,A} &\ = &\ F_i^{V,A}(
p_1 = -p_{\bar{t}}, \ p_2 = p_{\bar{b}} ,\ m_1 = m_t, \ m_2 = m_b ) \nonumber
\eea

\vspace{0.5cm}

The statndard definition for the (3-point) Passarino-Veltman functions is:
\bec{c_tensor}
{\cal{C}}^{ \{0,\mu,\mu \nu \} } =  
\int \frac{d^4 k}{i \pi^2}\ \frac{ \{1,k^\mu,k^\mu k^\nu \} }
{(k^2 - m_1^2)\ ((k+p_1)^2 - m_2^2)\ ((k+p_1+p_2)^2-m_3^2)}\
\eec
with the  scalar functions ${\cal{C}}_{ij}(p_1^2,p_2^2,m_1^2,m_2^2,m_3^2)$
 given by:
\bea
\label{c_decom}
{\cal{C}}^{\mu} & = & p_1^{\mu} {\cal{C}}_{11} + p_2^{\mu} {\cal{C}}_{12} \\
{\cal{C}}^{\mu \nu}& = & p_1^{\mu} p_1^{\nu} {\cal{C}}_{21} + 
p_2^{\mu} p_2^{\nu} {\cal{C}}_{22} + 
\left(p_1^{\mu} p_2^{\nu} + p_2^{\mu} p_1^{\nu}\right){\cal{C}}_{23}
+ g^{\mu \nu} {\cal{C}}_{24} \nonumber
\eea
The expressions for the reduction of  ${\cal{C}}_{ij}$ functions 
in terms of the scalar one-, two- and three-point
integrals ${\cal{A}}^0, {\cal{B}}^0 $
$ {\cal{C}}^0$ are standard, and we do not give them here 
(see \cite{thesis} or \cite{hagiwara}).

The $C$ function used in Eqs. (\ref{f_vect}), (\ref{f_axial}) are defined by:
\bec{cc_decom}
C^{ \{0,\mu,\mu \nu \} } =  
\int \frac{d^4 k}{i \pi^2}\ \frac{ \{1,k^\mu,k^\mu k^\nu \} }
{(k^2 + i\epsilon)\ ((k - p_1)^2 - m_1^2)\ ((k + p_2)^2-m_2^2)}\
\eec
\bea
C^{\mu} & = & -p_1^{\mu} C_{11} + (p_2-p_1)^{\mu} C_{12}  \\ 
C^{\mu \nu}& = & p_1^{\mu} p_1^{\nu} C_{21} + 
(p_2-p_1)^{\mu} (p_2-p_1)^{\nu} C_{22} - 
\left[p_1^{\mu} (p_2-p_1)^{\nu} + (p_2-p_1)^{\mu} p_1^{\nu}\right]C_{23}
\nonumber\\ 
 & & +\ g^{\mu \nu} C_{24} \nonumber
\eea
therefore:
\bec{a}
C_0,C_{ij} = {\cal{C}}^0, {\cal{C}}_{ij}(p_1^2,(p_2 - p_1)^2,0,m_1^2,m_2^2)
 .
\eec
Moreover, the function $B_0^{12}$ in Eq. (\ref{f_vect}) is given by:
\bec{jfdg}
B_0^{12} \ = \ {\cal{B}}^0((p_1+p_2)^2,m_1^2,m_2^2)
\eec 
and in Eqs. (\ref{f_fse}):
\bec{fhgkh}
B_n(p^2,m^2) = {\cal{B}}_n(p^2,0,m^2) \ , \ \hbox{for  } n = 0 , 1.
\eec



\end{document}